# Intrinsic Nano-scale Phase Separation of Bulk As$_2$S$_3$ Glass

D.G. Georgiev, P. Boolchand
Department of ECECS
University of Cincinnati
Cincinnati, Ohio, 45221-0030

and

K.A. Jackson
Department of Physics
Central Michigan University
Mt. Pleasant, Michigan, 48858

## Abstract

Raman scattering on bulk As$_x$S$_{1-x}$ glasses shows that vibrational modes of As$_4$S$_4$ monomer first appear near x = 0.38, and their concentration increases precipitously with x, suggesting that the stoichiometric glass (x = 0.40) is intrinsically phase separated into small As-rich (As$_4$S$_4$) and large S-rich clusters. Support for the Raman active vibrational modes of the orpiment-like and realgar-like nano-phases is provided by ab-initio Density Functional Theory calculations on appropriate clusters. Nanoscale phase separation provides a basis to understand the global maximum in the glass transition temperature, T$_g$ near x = 0.40, and the departure from Arrhenius T-activation of As$_2$S$_3$ melt viscosities.

# § 1. Introduction

Orpiment ($As_2S_3$) crystallizes in a layer-like structure with each layer consisting of twelve membered $As_2(S_{1/2})_3$ rings (Morimoto 1954) forming elements of medium-range-structure. Each ring consists of six pyramidally coordinated $As(S_{1/2})_3$ units with bridging S atoms. The *intra-layer* bonding interactions are covalent, while the *inter-layer* ones are van der Waals in character and weaker. The vibrational density of states in orpiment have been examined in Raman scattering (Zallen 1974, Kawamura *et al.* 1983). Interlayer optic modes exhibit at least an order of magnitude larger pressure shifts (mode Gruneisen) than intralayer ones, confirming the quasi 2-d molecular structure in vibrational spectroscopy.

In analogy to orpiment the molecular structure of stoichiometric $As_2S_3$ *glass* is widely regarded to consist of a network of S-bridging pyramidal $As(S_{1/2})_3$ units (Myers and Felty 1967). Several studies have emphasized that the similarity of structure goes beyond the local order to the medium-range order, i.e., the glass network also possesses a layer-like structure (Tanaka *et al.* 1985, Shimakawa *et al.* 1998, Taylor *et al.* 1998) although that view is not shared universally (Yang et al. 1987). Evidence for presence of a small but finite (< 2%) concentration of like-atom bonds (As-As, S-S) in the stoichiometric glass has emerged from Raman scattering (Wagner *et al.* 1998, Frumar *et al.* 1997) and [129]I Mössbauer spectroscopy (Zitkovsky 1989).

An issue of central importance concerns the distribution of these like-atom bonds in glass structure. A common view is to regard such bonds as "defects" in a *fully polymerized* network. The like-atom bonds then form at *random* in a nearly *chemically ordered* but *continuous* random network. An alternate view (Phillips *et al.* 1980) is to regard the As-As bonds and S-S bonds to form part of two distinct molecular clusters which bond to each other by weaker (van der Waals) forces. In the latter model, internal surfaces become an intrinsic feature of glass structure and the network is viewed to be *partially* polymerized, i.e., phase separated on a *molecular* or *nanoscale*.

Nanoscale phase separation effects are of general interest in glass science (Boolchand *et al.* 2002). Such structural effects produce usually pronounced changes in glass physical properties including a lowering of the glass transition temperature, a lowering of the optical band gap, an increase of molar volumes as network global connectivity diminishes. These pronounced changes in physical properties can mask the more subtle elastic effects related to rigidity transitions as discussed recently (Boolchand *et al.* 2002). In general, it appears that chalcogen-rich glasses are usually fully polymerized, but chalcogen-deficient ones tend to demix into characteristic nanophases although exceptions do occur. Stoichiometric glasses ($GeS_2$, $GeSe_2$, $As_2Se_3$, $As_2S_3$) that sit at the boundary between chalcogen-rich and chalcogen-deficient glasses have thus attracted much attention. The evidence (Bresser *et al.* 1981) suggests that the stociometric glasses of $GeSe_2$ and $GeS_2$ are composed of a majority phase and a cation-rich minority phase.

In this work, we present new Raman scattering and T-modulated DSC results on binary $As_xS_{1-x}$ glasses focusing near the stoichiometric composition x=2/5. The present results in conjunction with previous $^{129}$I Mössbauer effect results (Zitkovsky 1989) show that the stoichiometric $As_2S_3$ glass is *nanoscale phase separated* into small $As_4S_4$ molecules and compensating large S-rich clusters (Phillips *et al.* 1980). The finding is consistent with the global maximum of $T_g$ near the stoichiometric composition, and the non-Arrhenius T-activation of $As_2S_3$ liquid viscosities (Vinogradova *et al.*1967).

## §2. Experimental results

99.999% $As_2S_3$, elemental S and As lumps from Cerac, Inc., were used as starting materials to synthesize the S-rich (x < 0.40) and As-rich (x > 0.40) bulk $As_xS_{1-x}$ glasses. All weighings were performed in a pure and dry $N_2$ ambient. The starting materials sealed in evacuated (5 x $10^{-7}$ Torr) quartz tubings, were slowly reacted by heating to 700°C, and melts homogenized by continuous rocking in a furnace for 48 hours. Melt temperatures were then slowly lowered to $T_\ell$ + 50°C and melts equilibrated for several hours before a water quench. Glass transition temperatures $T_g(x)$ were measured using a model 2920 T-modulated DSC instrument from TA, Instruments, at a 3°C/min scan rate (heating followed by cooling) and a 1°C/100s modulation rate. $T_g(x)$ trends show an increase with As-content of the glasses at low x and a global maximum near x = 0.40 as shown in Fig. 1. The observed $T_g(x)$ trend closely parallels those of $T_\ell(x)$ (Fig. 1). Furthermore, $T_g(x)$ of a stoichiometric glass, melt quenched from $T_q$ = 550°C (open circle Fig. 1) is found to be 6°C lower than that of a sample melt quenched from $T_q$ = 350°C (filled circle Fig. 1). Raman scattering excited by 6mW of the 647.1nm radiation from a Kr-ion laser, focused to 50μm spot size, was studied in a back-scattering geometry. The scattered radiation was analyzed using a triple monochromator system (model T64000 from Instruments SA, Inc.) and a CCD detector.

Figure 2a shows Raman spectra of $As_2S_3$ glass samples synthesized at three different $T_q$s. Scattering strength of the 3 modes labeled R in the $150 < \bar{\nu} < 250$ cm$^{-1}$ range are found to steadily increase with $T_q$. Furthermore, in the main band centered near 340cm$^{-1}$, a feature near 360 cm$^{-1}$ becomes more conspicuous with increasing $T_q$. This group of modes labeled R in the glasses represent the Realgar-like nano-phase which consists of $As_4S_4$ molecules. Spectra of glasses near the composition x = 0.40 are shown in Fig. 2b. Of special interest is the R-mode at 190 cm$^{-1}$, which is not observed at x = 0.35 (see inset of Fig. 2b), but first appears near x = 0.38, and is found to monotonically grow in scattering strength with x thereafter.

Figure 3 compares Raman spectra of Orpiment and Realgar (β-$As_4S_4$) with those of glasses at x = 0.40 and 0.44. The Realgar sample was synthesized by annealing an $As_4S_4$ melt for 72 hours at $T_\ell$ - 25°C followed by an air quench, and the nature of the crystalline phase (β - $As_4S_4$) ascertained by powder XRD measurements. The Raman spectra of these crystalline phases are in excellent accord with earlier reports (Kobliska and Solin 1973, Ward 1968). Figure 3 reveals that the narrow set of modes labeled R and observed in Realgar have close parallels in the spectra of the glasses. On the other hand, the sharp modes seen in orpiment and labeled O, also have close parallels in the spectra of the glasses, except they are blue-shifted by about 18 cm$^{-1}$ and significantly broadened in the disordered phase.

## §3. Discussion

### 3.1. Raman mode assignments

Our interpretation of these Raman results is as follows. The set of sharp modes labeled R in the glasses are traced to the presence of As$_4$S$_4$ monomers in the glasses. These intra-cage bond-stretching mode frequencies in Realgar and those in the glasses are the same (within ± 5 cm$^{-1}$) because the cages are largely *decoupled* from the host networks. The R modes are broadened in glasses in relation to those in the crystal because of the presence of packing stresses in the disordered phase. The modes labeled O in the glasses are identified with pyramidal As(S$_{1/2}$)$_3$ units that probably form rings whose size is less certain. The 18 cm$^{-1}$ blue shift of the O-modes between orpiment and the glasses reflects a transfer of bonding strength from inter-layer (van der Waals) interactions to intra-layer (covalent) ones due to loss in planarity of the layers in the glasses. A parallel example occurs in Se, where the bond stretching mode at 233 cm$^{-1}$ in *trigonal Se* is known to upshift to 250 cm$^{-1}$ in *Se-glass* due to the more *molecular* nature of chains in the disordered phase (Zallen and Lucovsky 1976). The O-modes are broadened in glasses because As(S$_{1/2}$)$_3$ units are optimally constrained and can therefore distort, in sharp contrast to modes of Ge(S$_{1/2}$)$_4$ tetrahedra that continue to be narrow in glasses because of the overconstrained nature of tetrahedra, and therefore less amenable to distortion.

*3.2. Density functional theory calculations on clusters*

The strong similarity between the g-$As_{44}S_{56}$ and c-$As_4S_4$ spectra shown in Fig. 3 is compelling evidence for the formation of $As_4S_4$ monomer units in the glass. To further support this conclusion, we have performed first-priniciples calculations based on the density functional theory (DFT). Specifically, we investigate two cluster models that feature As-As bonds incorporated in different ways into the glass network. The first model features the cage-like $As_4S_4$ unit (Fig.4) that forms the basic building block of realgar. Here, the two As-As bonds are connected by four bridging S atoms to form a closed monomer unit that satisfies the valence requirements of all the atoms. In the second model, we open the monomer cage and add two more S atoms. The As dimers still share two bridging S atoms, but the bonds to the remaining S atoms mimic connections to the glass network. H atoms are used to maintain 2-fold coordination of the outer S atoms in this model. The electronegativity of H (2.1) is similar to that of As (2.0), so that no large, unphysical charge transfers are expected due to the S-H bonds.

The DFT calculations were carried out using a Gaussian-orbital-based methodology that has been discussed in detail elsewhere (Pederson and Jackson 1990). Pseudopotentials are used for the heavy atoms, while the H atoms are treated in an all-electron manner (Briley *et al.* 1998). The cluster models are first relaxed to a local minimum energy configuration. The full vibrational spectrum (including frequencies and mode eigenvectors) of each model is then calculated by diagonalizing the appropriate force constant matrix. Finally, the Raman activities associated with each vibrational mode are calculated directly using the DFT[20], allowing us to compare and contrast the Raman signature of the two models.

In Table 1 we present results for the two models shown in Fig. 4. The table lists all the modes in both clusters that have significant total Raman intensity in the frequency range 170 – 400 $cm^{-1}$. The Raman-active modes for both models can be grouped into a lower frequency set (180-230 $cm^{-1}$) and a higher frequency set (350-390 $cm^{-1}$). Analysis of the mode eigenvectors shows that the lower frequency modes correspond mainly to As-As bond stretches, with some admixture of As-S bond bending. The higher frequency modes correspond to As-S bond stretches.

The calculated Raman-active modes for the monomer model are fully consistent with the assignments made in Fig. 2 and Fig. 3. The lower frequency modes, at 183 and 221 $cm^{-1}$, are an excellent match to the positions of the strong, lower frequency peaks in the c-$As_2S_2$ spectrum and to the corresponding peaks the in g-$As_{44}S_{56}$ spectrum (185 and 220 $cm^{-1}$, respectively). In addition, the calculated intensities (13.1 vs. 6.7 for the 183 and 221 $cm^{-1}$ modes, respectively) agree qualitatively with the observed intensities. This is seen most clearly in the top two panels of Fig. 3. The higher frequency modes of the monomer model, at 371 and 355 $cm^{-1}$, are also in good agreement with strong Raman peaks in the crystal and glass spectra at 361 and 346 $cm^{-1}$, respectively. Again the calculated intensities (49.4 and 8.2 for the 371 and 355 $cm^{-1}$ modes, respectively) are in good agreement with the corresponding observed intensities. The fact that the calculated positions in both cases are stiffer by about 10 cm$^{-1}$ suggests that this shift is a systematic overestimate of the As-S bond stretch frequency by the DFT. To test this, we also computed the average stretch mode frequency for the AsS$_3$ pyramid, the basic building block of g-As$_2$S$_3$. It is reasonable to expect that the calculated stretch mode for the molecular model will be near the centroid of the Raman band of the glass. The calculated symmetric stretch frequency is 362 cm$^{-1}$, while the centroid in the observed spectrum is near 350 cm$^{-1}$. Thus the DFT frequency for As-S stretches appears indeed to be somewhat stiffer than observed.

We note that the c-As$_2$S$_2$ Raman spectrum shown in Fig. 3 includes a third high frequency peak near 350 cm$^{-1}$ that has no analogue among the Raman-active monomer frequencies shown in Table I. We suspect that this peak corresponds to a doublet of asymmetric As-S stretch modes at 343 cm$^{-1}$ in the isolated monomer spectrum. These modes have very little Raman intensity in the monomer model, but have strong IR intensity. Coupling between monomer units in the crystal may transfer additional Raman intensity to combinations of these modes, giving rise to the additional peak in the crystal spectrum.

The results given in Table 1 show clear differences between the monomer and network models. In the low frequency range, the network model has three Raman-active modes, at 195, 206 and 227 cm$^{-1}$, respectively, compared to two modes in the monomer model (at 183 and 221cm$^{-1}$). The network model also has an extra mode in the high frequency range, at 388 cm$^{-1}$, that is not present in the monomer spectrum. The remaining high frequency, Raman-active modes are close to corresponding modes in the monomer spectrum.

In contrast to the monomer model, the calculated Raman-active modes for the network model do not agree well with the positions and intensities of peaks observed in the c-$As_2S_2$ or g-$As_{44}S_{56}$ spectra shown in Fig. 3. This is particularly true in the low frequency part of the spectrum, where the network model has modes at 195 and 206 cm$^{-1}$. These are stiffer than the observed peak at 185 cm$^{-1}$. The calculated 227 cm$^{-1}$ mode is also stiffer than the observed peak at 220 cm$^{-1}$. The calculated relative intensities of the network model modes (6.4, 6.0 and 11.8, respectively for the 195, 206 and 227 cm$^{-1}$ modes) also disagree with the intensities of the observed peaks, which suggest that the lower frequency modes should be more intense. Interestingly, the calculated 227 cm$^{-1}$ mode lies near the observed peak at about 232 cm$^{-1}$ that appears in the g-$As_xS_{1-x}$ spectra, but not in the c-$As_2S_2$ spectrum. As seen in the inset in Fig. 2, the 232 cm$^{-1}$ peak appears at around x=0.35 and grows in intensity before the 221 cm$^{-1}$ peak appears at around x=0.4. A possible interpretation is that as x increases to 0.35 isolated As-As bonds are first formed in the glasses. The associated stretch modes are stiffer than those in monomer units and account for the observed Raman peak near 232 cm$^{-1}$. As the concentration of As-As bonds increases with x, $As_4S_4$ monomer units begin to form. The Raman signature of the monomers is clearly evident at x=0.40 and the monomer becomes the dominant structural motif for As-As bonds by around x=0.42.

### 3.3. Broken chemical order in $As_2S_3$ glass

To test these ideas further, we deconvoluted the observed Raman lineshape of the glasses in terms of a superposition of requisite number Gaussians keeping linewidths, centroids and intensities of each mode unrestricted. The result of the analysis appears in Fig. 5, in which the modes marked with arrows and labeled R represent those of $As_4S_4$ monomers while the remaining ones represent those of pyramidally coordinated $As(S_{1/2})_3$ units in orpiment-like sheets. The results of Fig. 5 show a systematic growth in scattering strength of R-modes at the expense of O-modes as As content of glasses increases in the 0.38 < x < 0.44 range. The deconvolution of the lineshape at x = 2/5, has largely been made possible because of a smooth extrapolation of the Raman results of the As-rich glasses that display sharper features. In the analysis of the broad band centered at 340 cm$^{-1}$, we found that the relative intensities of the Realgar-like modes remains relatively constant within error of measurement. The integrated area under the Realgar-like Raman modes normalized to the total area, $I_R/I(x)$ is found to increase with x almost linearly once x ≥ 0.38 and at x=2/5, the ratio $I_R/I(x)$ acquires a value of 0.16(5).

We make use of the Raman cross-sections of the R-like and pyramidal $As(S_{1/2})_3$ modes, given by the clusters DFT calculations, and obtain a fraction of 7% of homopolar As bonds in the $As_2S_3$ glass. The broken chemical order inferred from Raman measurements may be compared to a value of 1% established by chemical dissolution of the glass in KOH which selectively extracts As only from $As_4S_4$ units (Kosek *et al.* 1983). The difference between our calculation and the result reported in Ref. 21 is considerable and is certainly due to the oversimplified cluster models and the rather indirect procedure involved in the estimate of the Raman cross-sections.

On stoichiometric grounds, the presence of As-As bonds requires S-S bonds to occur in $As_2S_3$ glass. Evidence for S-S bonds in $As_2S_3$ glass emerged independently from $^{129}I$ Mössbauer spectroscopy (Zitkovsky 1989). Here one alloys $^{129m}Te$ tracer and infers the chemical environment of the chalcogen parent by examining the $^{129}I$ nuclear hyperfine structure of the former (Boolchand 2000). The experiments show a bimodal (A, B) site distribution (Frimar *et al.* 1997, Zitkovsky 1989) with $I_B/I_A = 1.60(5)$. Here, the oversized $^{129m}Te$ tracer segregates to *surfaces* of large S-rich orpiment-like clusters to *selectively populate* the roomy and chemically disordered S-S edge sites (Zitkovsky 1989, Phillips *et al.* 1980) (B) over the S-bridging cluster *interior* (A) sites. The result is a nearly 70-fold enhancement of the B over the A site population ($I_B > I_A$) which constitutes evidence for existence of internal surfaces dressing the S-rich cluster. In a homogeneous continuous random network model of $As_2S_3$ glass such a large population enhancement, $I_B > I_A$ would be precluded. Furthermore, the present Raman results (Fig. 2a) show that the degree of nanoscale phase separation in $As_2S_3$ glass increases with $T_q$ as also noted in a previous study (Mamedov *et al.* 1998). The free energy of the phase separated liquid ($F = U - TS + PV$) is apparently lowered by the creation of internal surfaces that provide an additional source of enthalpy (U), i.e., van der Waals bonding and entropy (S) particularly if the clusters are small and can move, to overwhelm the endothermic reaction (Pauling 1960)

$$2 (As - S) \rightarrow (As - As) + (S - S) - 11.5 \text{ kcal/mole} \qquad (1)$$

and promote phase separation as $T_q$ is increased.

Unlike Raman scattering, $^{75}$As NQR lacks the sensitivity (Taylor *et al.* 1998) to detect the < 2% concentration of As-As bonds in $As_2S_3$ glass. Furthermore, since the size of the S-rich clusters is the order of visible light wavelength (as revealed by absorption band tails (Tanaka 2001)), NQR can only probe the *majority* (> 98%) As sites in the S-rich cluster interior, which are found to be orpiment-like as confirmed by the measured NQR parameters (Taylor *et al.* 1998).

### *3.4. Broken chemical order and $T_g$ maximum*

The Pauling single bond-strength (Pauling 1960) of a S-S bond of 50.9 kcal/mole *exceeds* that of an As-S bond of 47.25 kcal/mole. The increase of $T_g$ upon alloying As in a sulfur glass is thus *qualitatively incompatible* with chemical bond strength arguments (Tichy and Ticha 1995) that would require $T_g(x)$ to decrease with x in the present case. $T_g$ reflects *global connectivity* of a glass network with differences in covalent bond strengths playing little role, if any, as elucidated by stochastic agglomeration theory (Micoulaut 1998) (SAT). According to this theory the increase of $T_g$ upon alloying As or Ge in a Se-glass reflects the increased cross-linking or connectivity of the alloyed network. The global maximum in $T_g$ and also of $T_\ell$ at the stoichiometric compositions in $GeSe_2$ (Bresser *et al.* 1981) and $As_2Se_3$ (Georgiev *et al.* 2000) glasses is a direct consequence of *nanoscale phase separation of the glass and the liquid*, respectively. In these systems , nano-scale phase separation effects are traced respectively to Ge-Ge bonds and As-As bonds not forming part of the backbone once the cation concentration approaches the stoichiometric composition, but nucleating separate clusters as revealed by Raman and Mossbauer spectroscopic measurements. The lowering of $T_\ell$ for off-stoichiometric compositions is a natural consequence of the Born elastic shear instability (Born 1939) due to presence of internal surfaces in corresponding crystals.

The reduction in $T_g$ of an $As_2S_3$ glass with increasing $T_q$, reflects the loss in global connectivity as the concentration of $As_4S_4$ molecules increases (Fig. 2a) and internal surfaces emerge. In Fig. 1, the sign change in the slope $dT_g/dx$, from a positive value at $x = 0.35$ to a negative value at $x = 0.42$ reflects the bonding role of As atoms, which serve to increase cross-linking of a S-rich ($x = 0.35$) network with the free S, but to depolymerize the network of an As-rich glass ($x = 0.42$) by clustering it. The present results thus form part of a general trend noted in the chalcogenides and is discussed in more details in recent reviews (Boolchand *et al.* 2002, Bresser *et al.* 1981).

### 3.5. *Broken chemical order and non-Arrhenius T-variation of viscosity*

A count of Lagrangian bonding constraints (Georgiev *et al.* 2000) shows that a network of pyramidally coordinated $As(S_{1/2})_3$ units is optimally constrained, i.e., 3 constraints/atom. $SiO_2$ glass, composed of a network of tetrahedral building blocks is also optimally constrained, i.e. $n_c=3$, because the bond-bending constraints associated with bridging oxygen are intrinsically broken (Zhang and Boolchand 1994). If $As_2S_3$ glass in analogy to the *optimally* (Zhang and Boolchand 1994) constrained oxide glasses ($SiO_2$ and $GeO_2$) were to also form a *continuous random network*, liquid viscosities would display a strong behavior, i.e. an Arrhenius activation with a low activation energy. Viscosities of $As_2S_3$ and $As_2Se_3$ have been measured (Phillips *et al.* 1980) and in sharp contrast to those of $SiO_2$ or $GeO_2$ melts *show* an Arrhenius T-activation with a higher activation energy than in $SiO_2$ and $GeO_2$ glass. The results show that both $As_2S_3$ and $As_2Se_3$ are mildly fragile. The nanoscale phase separation of the $As_2S_3$ network contributes fragility (Angell 2000) that results in the observed deviation from strong behavior for much the same reason that addition of alkali to $SiO_2$ disrupts the continuous random network and renders it fragile.

## §4. Conclusions

Raman lineshape deconvolution of $As_{0.4+y}S_{0.6-y}$ glasses at y = 0.04, when extended to y = 0, show that $As_2S_3$ glass is nanoscale phase separated into small As-rich and large S-rich clusters. These results provide a structure basis to understand the local maximum in $T_g$, the broken chemical order inferred in earlier $^{129}$I Mössbauer spectroscopy measurements, weak optical absorption tails, the non-Arrhenius T-activation of $As_2S_3$ liquid viscosities, and are parallel to results found on other stoichiometric chalcogenide glasses.

## Acknowledgements

We have benefited from discussions with D. McDaniel. The orpiment sample used in the present work was provided by R. Zallen. The assistance of J.Spijkerman and J.Thayer in the course of experiments is acknowledged with pleasure. The work at University of Cincinnati was supported by NSF grant DMR 01-01808.

Table 1. Raman-active modes for two cluster models featuring As-As bonds. The first column identifies the model (see Fig. 4). The second column gives the frequencies of the Raman-active modes. The third and fourth columns give the total Raman intensities and depolarization ratios ($I^{HV}/I^{VV}$) of the modes.

| Cluster | $\omega(cm^{-1})$ | $I^{Ram}$ (Å$^4$/amu) | $\rho$ |
|---|---|---|---|
| Monomer | 183 | 13.1 | 0.16 |
|  | 221 | 6.7 | 0.06 |
|  | 355 | 8.2 | 0.75 |
|  | 371 | 49.4 | 0.01 |
| Network | 195 | 6.4 | 0.14 |
|  | 206 | 6.0 | 0.03 |
|  | 227 | 11.8 | 0.24 |
|  | 351 | 18.0 | 0.33 |
|  | 369 | 32.1 | 0.02 |
|  | 371 | 33.2 | 0.08 |
|  | 388 | 6.9 | 0.75 |

Figure Captions

**Fig. 1**. Compositional trends in glass transition temperature ($T_g$) and liquidus temperature ($T_l$) showing a global maximum near x = 0.40 in the titled binary. Inset shows MDSC traces of the reversing heat flow and the non-reversing heat flow of a sample at x = 0.10 with $T_g$ = 38°C and the Sulfur polymerization transition temperature, $T_\lambda$, near 140°C.

**Fig.2.a**. Raman scattering of $As_2S_3$ glass samples quenched from indicated melt temperatures ($T_q$) showing growth in scattering of the modes labeled R with $T_q$. Inset shows the reduction in $T_g$ from 214°C to 206°C when $T_q$ is increased from 350°C to 550°C.
**b.** Raman scattering in binary $As_xS_{1-x}$ glasses at indicated x showing onset of the R modes at x = 0.38. The inset shows a magnified view of the R modes. The mode at 185 $cm^{-1}$ at the arrow location onsets near x =0.38.

**Fig.3**. Raman scattering in the two crystalline phases, Orpiment (c-$As_2S_3$) and Realgar (β-$As_4S_4$) compared to the lineshapes observed in the two indicated glasses at x = 0.44 and 0.40. Features labeled R in the glasses are identified with modes of Realgar, while features labeled O in the glasses are identified with modes of Orpiment. The orpiment modes in the glasses are blue-shifted by 18 $cm^{-1}$ and broadened in relation to the crystalline phase. See text for details.

**Fig. 4.** Cluster models of $As_4S_4$ in a monomeric and polymeric form. The dark/light atoms are As/S. Hydrogen atoms are used to maintain two-fold coordination of S atoms in the network model.

**Fig. 5**. Raman lineshape deconvolution of the observed spectra (points) in terms of a superposition of Gaussians composed of modes of Realgar (arrows) and orpiment with no restrictions on the centroid, width and intensity of each Gaussian. Note that the Realgar mode scattering strengths increase with x.

Yang, C.Y., Sayers, D.E., Paesler, M., 1987, Phys. Rev. B, **36**, 8122 .

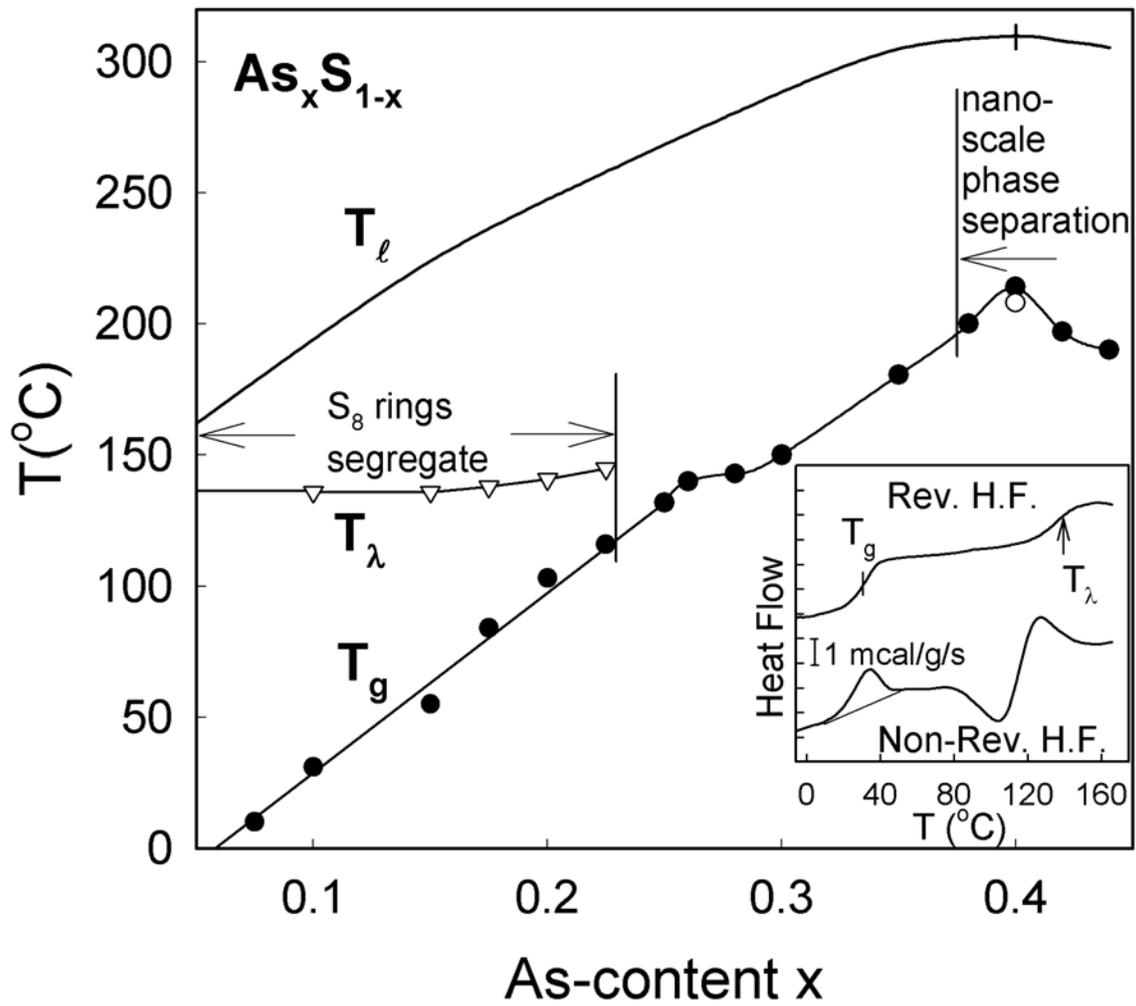

Fig.1

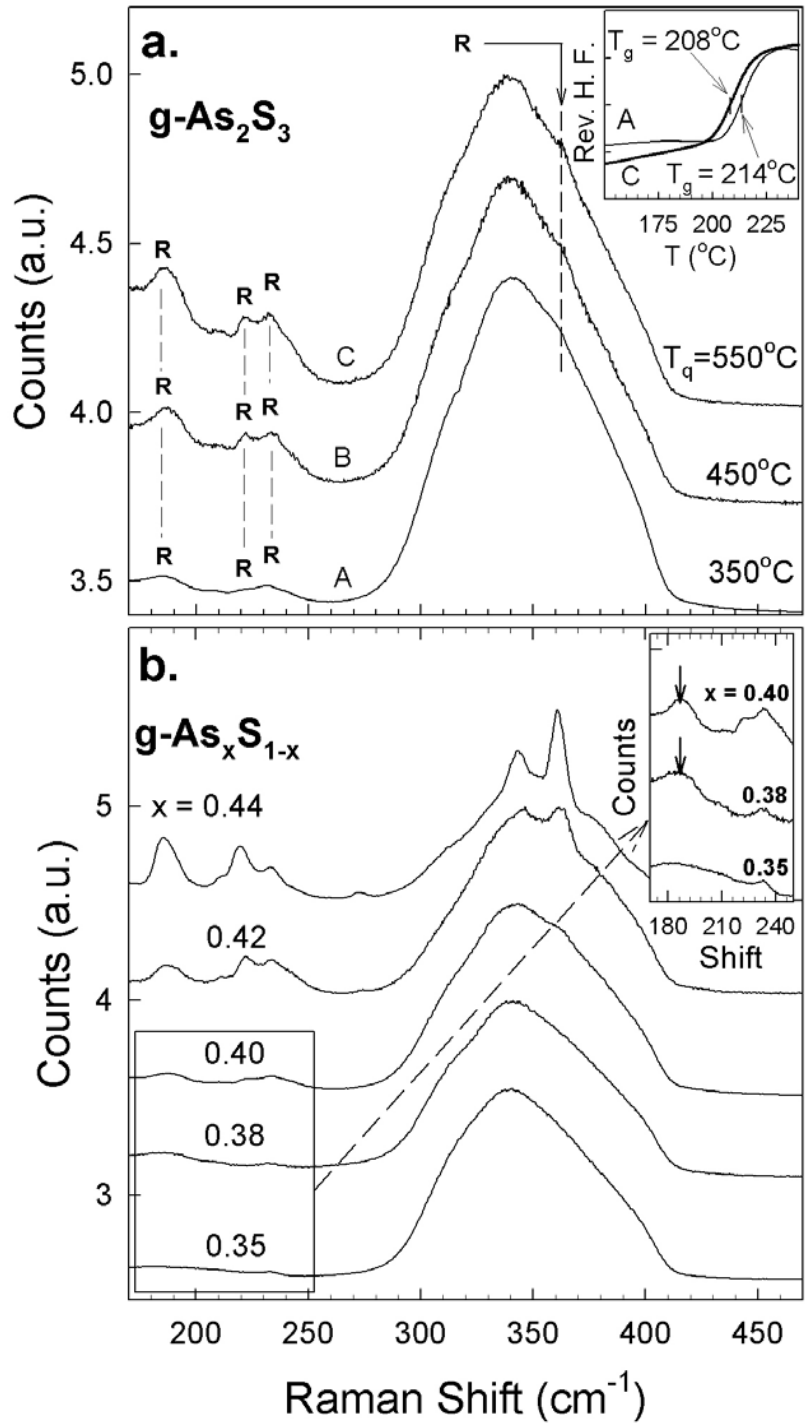

Fig.2

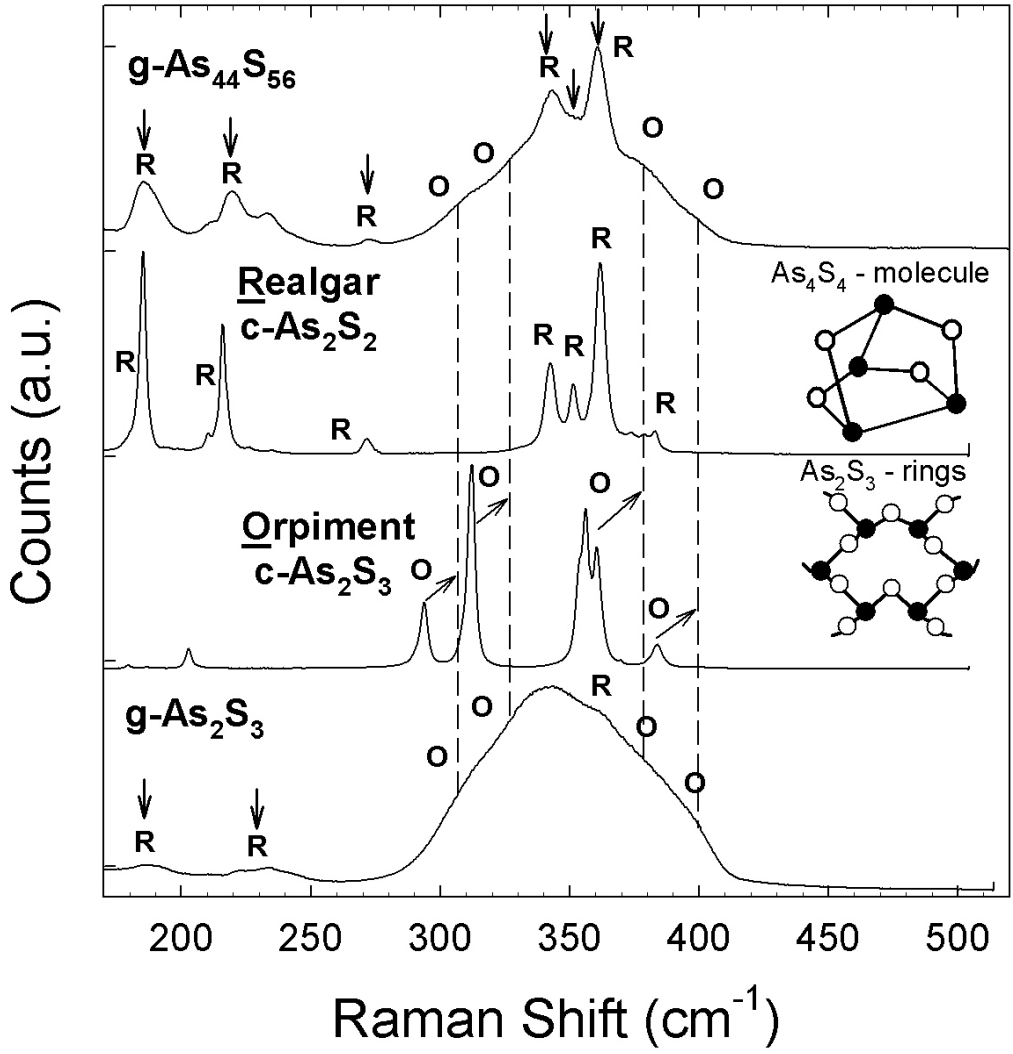

Fig.3

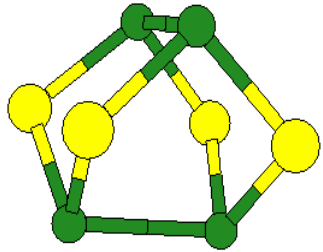

**Monomer**

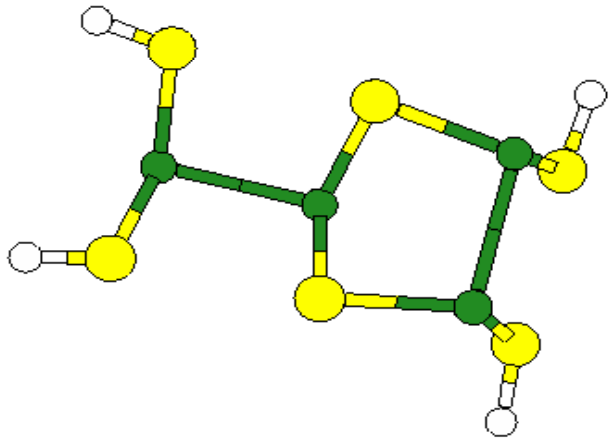

**Network**

Fig.4

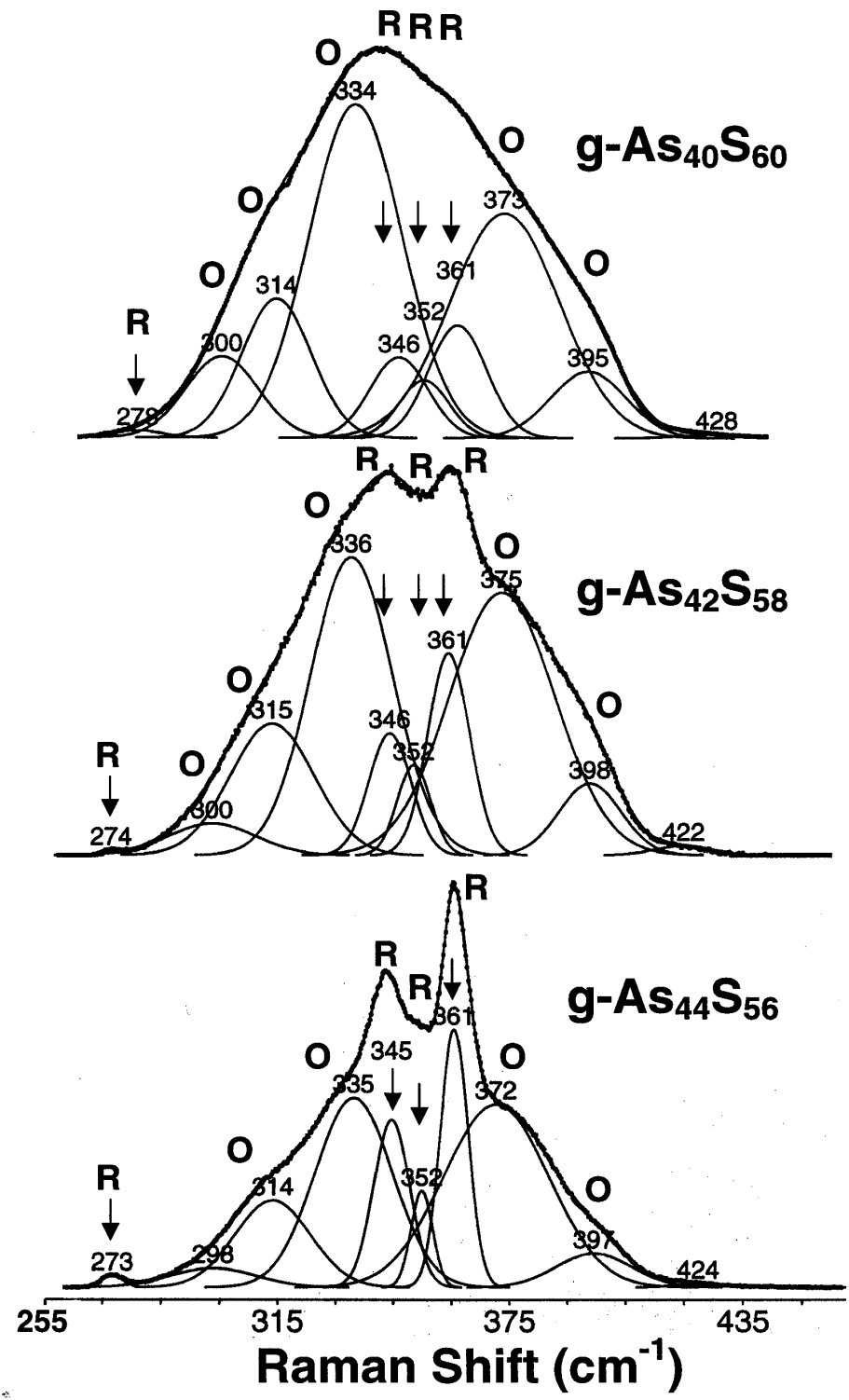

Fig.5